\documentclass{jps-cp}
\usepackage{txfonts} %Please comment out this line unless the txfonts package is availabe in your LaTeX system.

\usepackage{bm}
\usepackage{amsmath,amssymb}
\usepackage{color}

\def\newblock{\hskip .11em plus .33em minus .07em}

\title{Structure of Bound States with Coulomb plus Short-range Interaction}

\author{Chisato \textsc{Uno}$^{1}$ and Tetsuo \textsc{Hyodo}$^{2}$}

\inst{$^{1}$Department of Physics, Tokyo Metropolitan University, Hachioji 192-0397, Japan \\
$^{2}$Research Center for Nuclear Physics, The University of Osaka, Ibaraki, Osaka 567-0047, Japan}

\email{hyodo@rcnp.osaka-u.ac.jp}

\abst{
We study the structure of bound states appearing in systems governed by the Coulomb and short-range interactions. We analyze the binding energies and wave functions of the bound states generated by the Coulomb plus short-range potential. We demonstrate that Coulomb-induced shifts of the binding energy are closely correlated with the spatial distribution of the wave function. Furthermore, we show that the asymptotic behavior of wave functions of weakly bound states is qualitatively altered by Coulomb repulsion, leading to a modification of the near-threshold mass scaling that is otherwise universal for short-range interactions.}

\kword{Exotic hadrons, Coulomb interaction, Near-threshold states}

\begin{document}
\maketitle

%%%%%%%%%%%%%%%%%%%%
\section{Introduction}
\label{sec:intro}
%%%%%%%%%%%%%%%%%%%%

% exotic hadrons, hadronic molecule
In recent years, a number of exotic hadron candidates have been discovered~\cite{ParticleDataGroup:2024cfk}, and understanding their internal composition has become one of the central issues in hadron physics. These exotic states are regarded as new forms of hadronic matter beyond ordinary mesons ($q\bar q$) and baryons ($qqq$), and various interpretations for their structures, such as multiquark configurations and hadronic molecules, have been proposed~\cite{Guo:2017jvc,Brambilla:2019esw}.

% Coulomb interactions
Among these, hadronic molecular states are considered to be formed by the hadron-hadron interaction as the driving force. While this interaction mainly originates from the strong force, charged hadron pairs are also subject to the Coulomb interaction. Since the electromagnetic force is typically about two orders of magnitude weaker than the strong interaction, its effects are usually negligible. For instance, the binding energy of a kaonic nucleus is typically several tens of MeV, whereas that of a kaonic atom, bound primarily by the Coulomb attraction, is of the order of keV~\cite{Ohnishi:2017uni,Hoshino:2017mty,Hyodo:2020czb,Hyodo:2022xhp}. This clear separation of energy scales justifies the conventional neglect of Coulomb effects in hadron physics.

% exceptional hadrons
Recently, however, several near-threshold exotic hadrons have been found to possess binding energies comparable to those generated by the Coulomb interaction. Prominent examples are the $X(3872)$, with a binding energy of about 40 keV, and the $T_{cc}$, with about 360 keV~\cite{ParticleDataGroup:2024cfk,LHCb:2021vvq,LHCb:2021auc}. For such shallow bound systems, even a weak Coulomb effect would alter their qualitative nature. Indeed, lattice QCD studies of the $\Omega_{ccc}^{++}\Omega_{ccc}^{++}$ system indicate that the strong interaction alone produces a bound state, but the inclusion of Coulomb repulsion dissolves it into a resonance~\cite{Lyu:2021qsh}. A similar trend appears in nuclear physics: the ground state of $^{8}$Be, a well-known $\alpha$-cluster nucleus consisting of two $\alpha$ particles~\cite{Wiringa:2000gb}, becomes unbound once the Coulomb repulsion is considered~\cite{Braaten:2004rn,Higa:2008dn}. Conversely, the Coulomb attraction can generate new bound states, as discussed for the $\Xi^{-}$-$\alpha$ system, which forms a Coulomb-assisted bound state otherwise unbound by the strong force~\cite{Hiyama:2022jqh,Kamiya:2024diw}.

% this work
In this work, we investigate the influence of the Coulomb interaction on near-threshold bound states using a potential model that incorporates both Coulomb and short-range  forces. The two-body systems with a Coulomb plus short-range interaction has been studied in various context of the physics~\cite{Domcke,Kong:1998sx,Higa:2008dn,Mochizuki:2024dbf,Kinugawa:2025kqr}. For weakly bound $s$-wave systems with large scattering lengths, low-energy universality~\cite{Braaten:2004rn,Naidon:2016dpf} and characteristic mass scaling~\cite{Hyodo:2014bda,Hanhart:2014ssa} are known to emerge. The present study explores how these universal properties are modified by the presence of the Coulomb interaction.

%%%%%%%%%%%%%%%%%%%%
\section{Formulation}
\label{sec:formulation}
%%%%%%%%%%%%%%%%%%%%

%==========================
\subsection{Coulomb plus short-range potential}
\label{subsec:Cps}

% Schroedinger equation
We consider a nonrelativistic two-body system with reduced mass $\mu$, interacting through a spherically symmetric potential $V(r)$. For the $s$-wave ($\ell=0$) state, the wave function can be written as $\psi(\bm{r}) = u(r)/(\sqrt{4\pi}\, r)$, where the radial wave function $u(r)$ satisfies the Schr\"odinger equation
\begin{align}
	\left[-\frac{1}{2\mu}\frac{d^{2}}{dr^{2}}+V(r)\right]u(r)
	& =
	Eu(r) .
	\label{eq:Schroedinger}
\end{align}

% potential
We model the short-range strong interaction by a square-well potential with strength $V_{0}$ and range $b$, acting on top of the Coulomb potential:
\begin{align}
	V(r)
	& =V_{0}\Theta(b-r)+\frac{\alpha Z_{1}Z_{2}}{r} 
	\label{eq:potential}
\end{align}
where $\alpha \simeq 1/137$ is the fine-structure constant, and $Z_{1}$ and $Z_{2}$ denote the electric charges of the two particles. A negative (positive) value of $V_{0}$ represents an attractive (repulsive) short-range interaction. The Coulomb interaction is attractive (repulsive) for opposite (same) charges, corresponding to $Z_{1}Z_{2}<0$ ($Z_{1}Z_{2}>0$), respectively. In the limit $V_{0}\to 0$, the potential~\eqref{eq:potential} reduces to the pure Coulomb interaction, while setting $Z_{1}Z_{2}=0$ corresponds to a purely short-range potential.

%==========================
\subsection{Bound states}
\label{subsec:BS}

% bound state, radius
If the interaction is sufficiently attractive, bound-state solutions satisfying the boundary conditions $u(r\to 0)=u(r\to \infty)=0$ exist with $E=-B<0$. For an attractive square-well potential without the Coulomb interaction ($Z_{1}Z_{2}=0$, $V_{0}<0$), the binding energy $B$ is determined by the condition
\begin{align}
    \sqrt{-2\mu(V_0 + B)}\, \cot\bigl(\sqrt{-2\mu(V_0 + B)}\, b\bigr)
    + \sqrt{2\mu B}
    =0,
    \label{eq:fB_def}
\end{align}
which admits either no solution or a finite number of solutions depending on $V_{0}$ and $b$.
In the case of a purely attractive Coulomb interaction ($Z_{1}Z_{2}<0$, $V_{0}=0$), there always exist infinitely many bound states, and the binding energies are given by the Bohr radius $a_{B}=1/(\mu\alpha |Z_{1}Z_{2}|)$ as
\begin{align}
   B_n &= \frac{1}{2\mu a_{B}^2}\frac{1}{n^2}, \quad n=1,2,3,\ldots .
   \label{eq:BCoulomb}
\end{align}

% radius
The radial density distribution of a bound state is given by $|u(r)|^{2}$ in terms of the radial wave function. Accordingly, the mean-squared radius of the bound state is calculated as
\begin{align}
	\langle r^{2}\rangle
	& =
	\int_{0}^{\infty} dr\ r^{2}|u(r)|^{2} .
\end{align}
In the case of a purely attractive Coulomb interaction, the wave functions are known analytically. 
As a result, for example, the mean-square radii of the states are obtained as $\langle r^2\rangle=3a_B^{2}\ (n=1), \ 42a_B^{2}\ (n=2), \ 207a_B^{2}\ (n=3)$, and so on.

%%%%%%%%%%%%%%%%%%%%
\section{Results}
\label{sec:results}
%%%%%%%%%%%%%%%%%%%%

In the following, we present the properties of bound states by the Coulomb plus short-range potential~\eqref{eq:potential}. With the $^{8}{\rm Be}$ nucleus in mind, the system is modeled as an $\alpha\alpha$ two-body system, where the reduced mass is fixed to 1863~MeV and the interaction range is set to $b=1.5$~fm. By varying the charge product $Z_{1}Z_{2}$ and the depth of the attractive square-well potential $V_{0}\leq 0$, we investigate how the properties of the resulting bound states change. In Sec.~\ref{subsec:binding}, we clarify the relation between the binding energy and the wave function when the Coulomb interaction is introduced. In Sec.~\ref{subsec:weakbound}, we focus on weakly bound states and examine the effects of the Coulomb interaction on the wave-function properties and on the mass scaling behavior near the threshold.

%==========================
\subsection{Binding energies}
\label{subsec:binding}

% single bound state
We first consider the case without the Coulomb interaction ($Z_{1}Z_{2}=0$). For a potential depth of $V_{0}=-25$~MeV, a single bound state is obtained with a binding energy of $B\simeq 5.3$~MeV. When a repulsive Coulomb interaction with $Z_{1}Z_{2}=+4$ is included, the binding energy is reduced to $B\simeq 0.45$~MeV. In contrast, the inclusion of an attractive Coulomb interaction with $Z_{1}Z_{2}=-4$ increases the binding energy to $B\simeq 11$~MeV. These results clearly demonstrate that the binding energy is shifted depending on the sign of the Coulomb interaction. In particular, when the Coulomb repulsion overcomes the attractive strong interaction, the bound state is expected to turn into a resonance. Indeed, the bound state obtained for $V_{0}=-20$~MeV disappears once the repulsive Coulomb interaction with $Z_{1}Z_{2}=+4$ is introduced, indicating a transition from a bound state to a resonance. The resulting binding energies for $V_{0}=-25$~MeV are summarized in Table~\ref{tab:single}.

\begin{table}[tb]
\caption{Binding energies obtained with the Coulomb plus square-well potential.}
\label{tab:single}
\centering
\begin{tabular}{lllll}
\hline
$Z_{1}Z_{2}=0$       & $0$   & $+4$   & $-4$  & $-4$ \\
$V_{0}$ [MeV]        & $-25$ & $-25$  & $-25$ & $0$ \\
\hline
Second excited state & -     & -      & 0.085 & 0.199     \\
First excited state  & -     & -      & 0.389 & 0.794     \\
Ground state         & 5.3   & $0.45$ & 11    & -         \\
\hline
\end{tabular}
\end{table}

% level repulsion
The pure Coulomb attraction ($Z_{1}Z_{2}=-4$, $V_{0}=0$~MeV) leads to an infinite number of bound states given by Eq.~\eqref{eq:BCoulomb}. With the present parameter set, the binding energy of the ground state ($n=1$) is $0.794$~MeV. When an additional short-range attractive interaction with $V_{0}=-25$~MeV is introduced, the binding energy is reduced to $0.389$~MeV. This reduction can be interpreted as an energy shift caused by level repulsion with the bound state generated by the short-range interaction. A similar mechanism is known in the $K^{-}p$ system, where the 1s level of kanoic hydrogen is shifted repulsively due to the presence of the (quasi-)bound state $\Lambda(1405)$ generated by the strong interaction, as confirmed both theoretically and experimentally\cite{SIDDHARTA:2011dsy,Hoshino:2017mty,Hyodo:2022xhp}. 

% two bound states, results
Next, we consider deeper attractive short-range potentials for which two bound states, namely the ground and excited states, are formed. In order to quantify the effect of the Coulomb interaction, we denote the binding energy obtained with only the short-range interaction as $B_{0}$, that obtained with Coulomb plus short-range interactions as $B$, and the Coulomb-induced shift of the binding energy as $\Delta B = B-B_{0}$. A positive (negative) value of $\Delta B$ indicates an increase (decrease) in the binding energy due to the Coulomb interaction. 
The results for $V_{0}=-125$~MeV and $-250$~MeV with the Coulomb interaction $Z_{1}Z_{2}=\pm4$ are summarized in Table~\ref{tab:double}. 
As in the case shown in Table~\ref{tab:single}, we find that the binding energy increases (decreases) for all energy levels when an attractive (repulsive) Coulomb interaction is introduced, corresponding to $\Delta B>0$ ($\Delta B<0$).

\begin{table}[tb]
\caption{Binding energies and Coulomb-induced shifts obtained with the Coulomb plus square-well potential for $V_{0}=-125$~MeV and $V_{0}=-250$~MeV.}
\label{tab:double}
\centering
\begin{tabular}{lllllll}
\hline
$V_{0}$ [MeV] & level & $Z_{1}Z_{2}=0$ & \multicolumn{2}{l}{$Z_{1}Z_{2}=+4$} & $Z_{1}Z_{2}=-4$ \\
&  & $B_{0}$ [MeV] & $B$ [MeV] & $\Delta B$ [MeV] & $B$ [MeV] & $\Delta B$ [MeV] \\
\hline
$-125$ & excited state & 9.55    & 2.06    & $-7.5$ & 17.1  & 7.6    \\
$-125$ & ground state  & 93.3    & 85.72   & $-7.6$ & 101   & 7.7   \\
$-250$ & excited state & 112.5   & 102.1   & $-10.0$ & 121.7 & 9.2     \\
$-250$ & ground state  & 214.8   & 206.7   & $-8.1$ & 222.9 & 8.1     \\
\hline
\end{tabular}
\end{table}

\begin{figure}[tb]
\centering
\includegraphics[width=0.4\textwidth,bb=0 0 600 400]{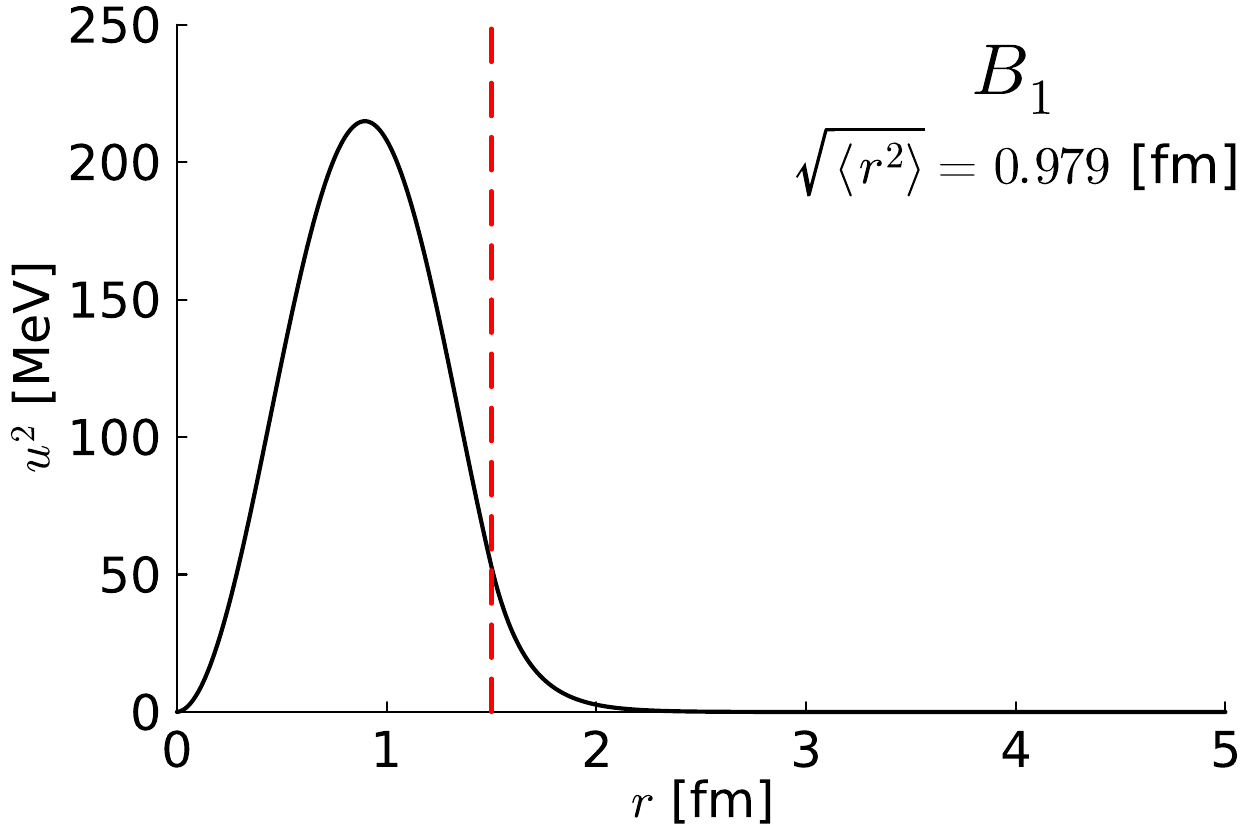}
\includegraphics[width=0.4\textwidth,bb=0 0 600 400]{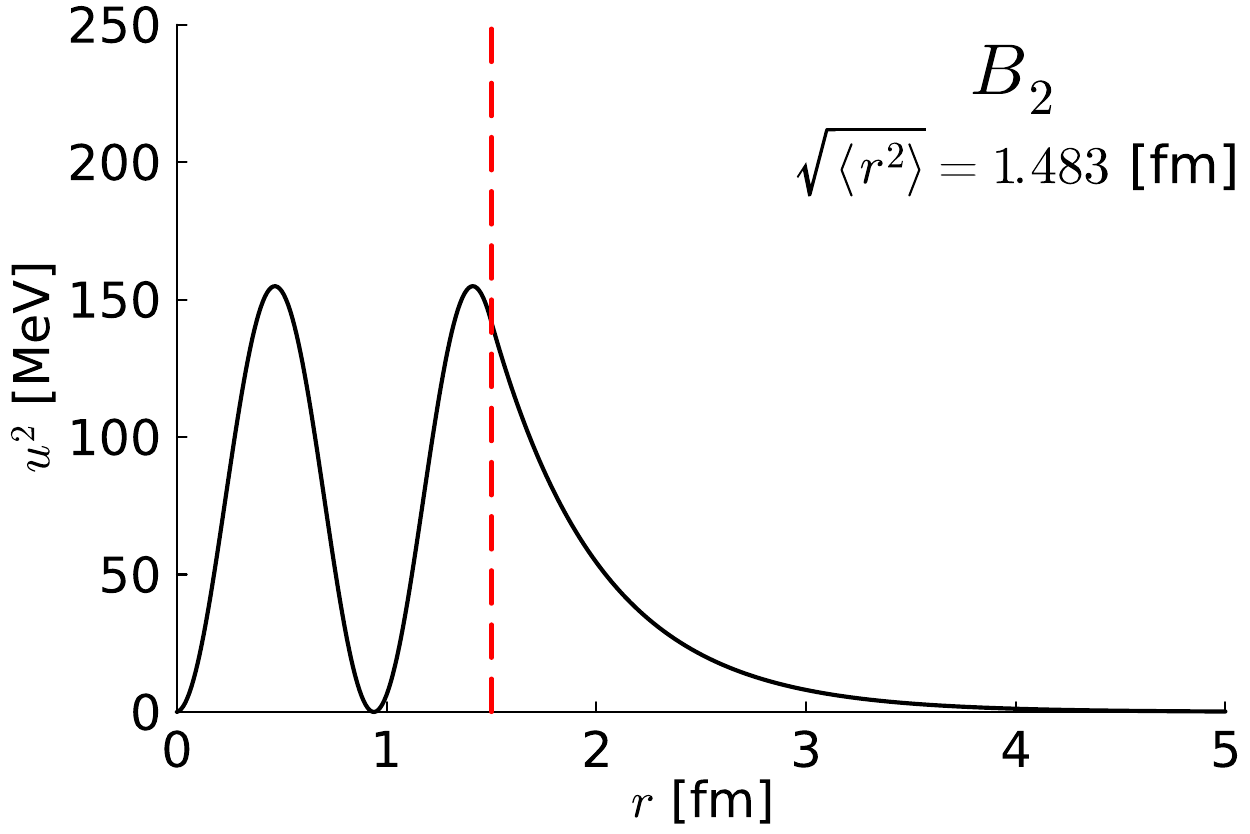}
\includegraphics[width=0.4\textwidth,bb=0 0 600 400]{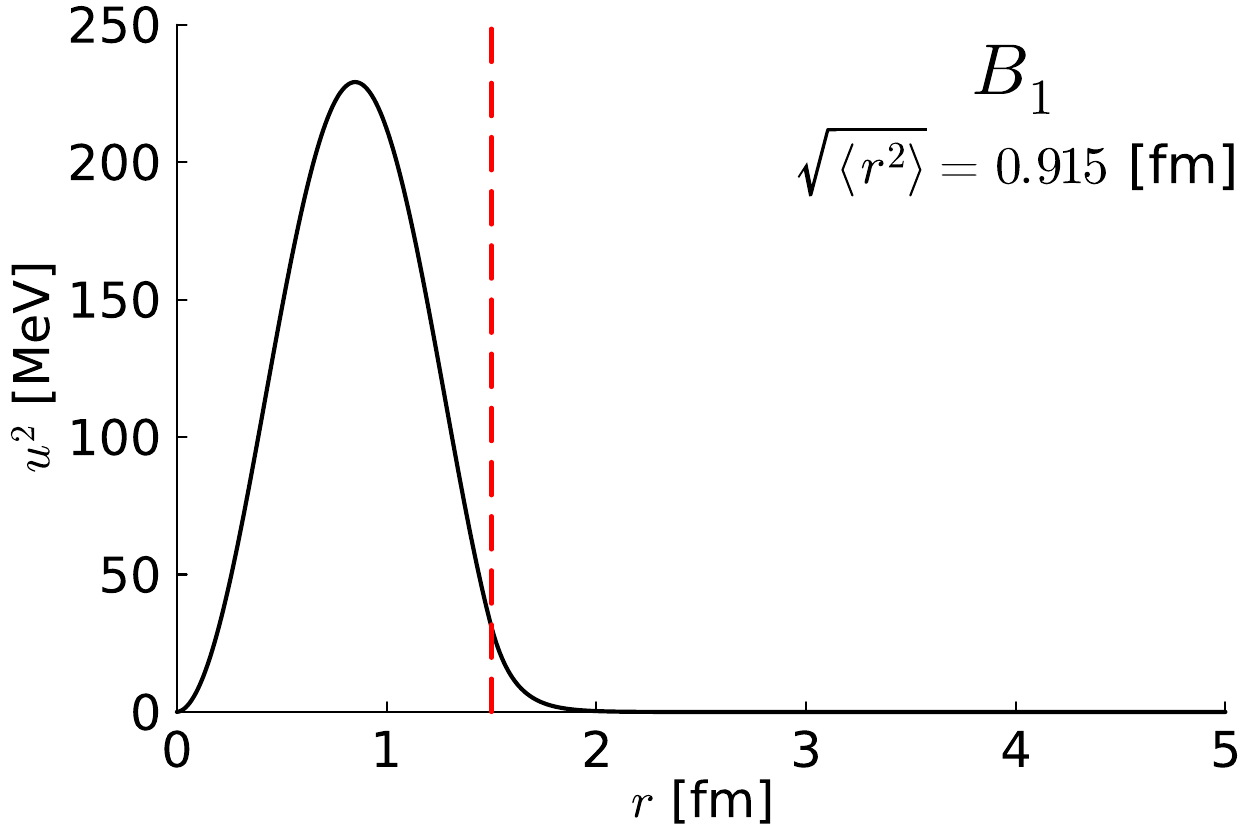}
\includegraphics[width=0.4\textwidth,bb=0 0 600 400]{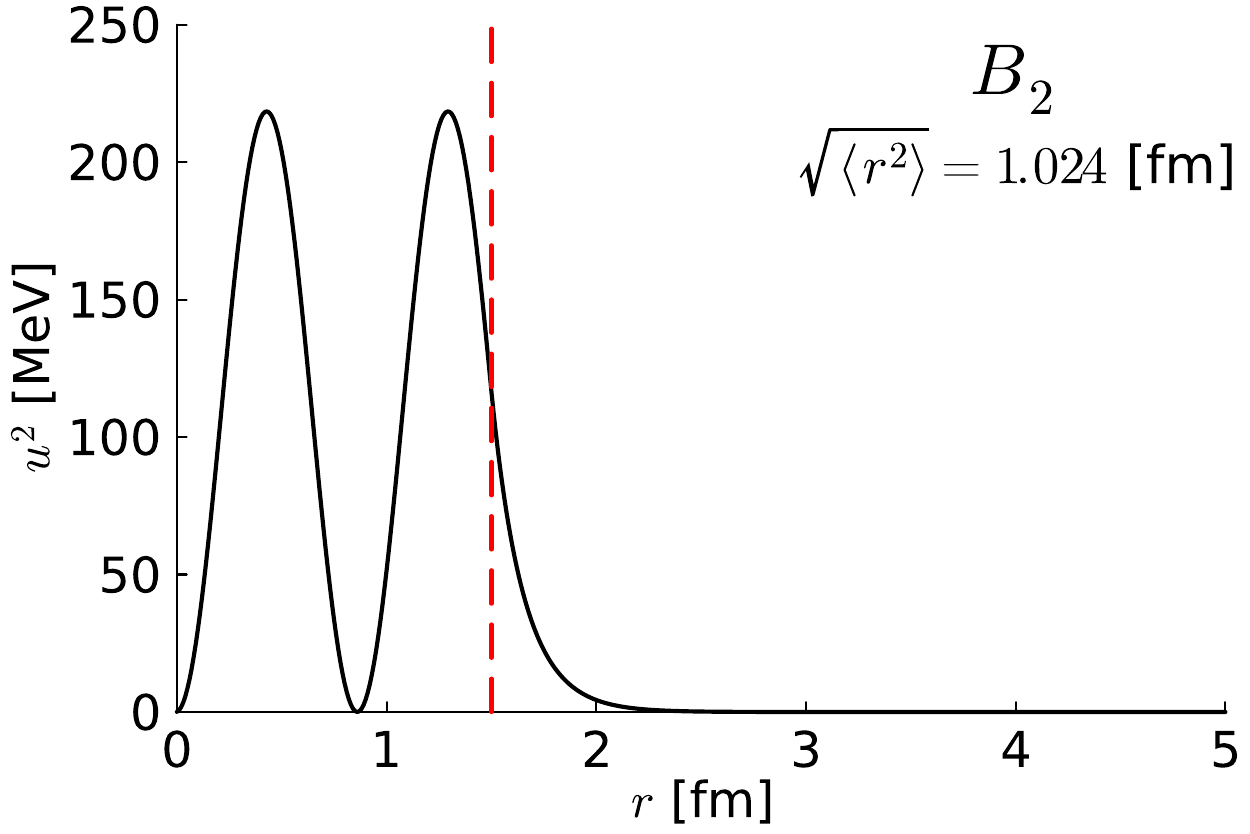}
\caption{Radial density distributions of bound states, $|u(r)|^{2}$, obtained with the square-well potential. Left (right) panels show the ground (excited) state. Top (bottom) panels correspond to $V_{0}=-125$~MeV ($-250$~MeV). The vertical dashed lines indicate the interaction range $r=b$.}
\label{fig:wf}
\end{figure}

% ground state and excited state
On the other hand, when we compare the magnitudes of the energy shifts $\Delta B$ for the ground and excited states, we find that their ordering depends on the depth of the potential well. Denoting the Coulomb-induced shifts of the ground and excited states as $\Delta B_{1}$ and $\Delta B_{2}$, respectively, we observe that $|\Delta B_{1}|<|\Delta B_{2}|$ for $V_{0}=-125$~MeV, whereas the opposite relation, $|\Delta B_{1}|>|\Delta B_{2}|$, holds for $V_{0}=-250$~MeV. 

% deep case
This difference can be understood from the bound-state wave functions before introducing the Coulomb potential. Figure~\ref{fig:wf} shows the density distributions of the bound states with $Z_{1}Z_{2}=0$. For $V_{0}=-250$~MeV (bottom panels), both the ground state (left, $B_{0}\sim215$~MeV) and the excited state (right, $B_{0}\sim113$~MeV) have wave functions localized within the interaction range, $r=b$. In this case, the excited-state wave function possesses a node at $r\sim0.8$~fm, which enhances the density near the origin where the Coulomb interaction is strongest, compared with the ground state. As a result, the excited state experiences a larger Coulomb-induced energy shift.

% shallow case
In contrast, for $V_{0}=-125$~MeV (top panels), the excited state (right) has a much smaller binding energy, $B_{0}\sim9.6$~MeV, and its density distribution extends well beyond the interaction range. Indeed, calculating the root-mean-square radius, we find that while all other states have a typical size of $\sqrt{\langle r^{2}\rangle}\sim 1.0$~fm, only the excited state for $V_{0}=-125$~MeV has a significantly larger radius of  $\sqrt{\langle r^{2}\rangle}\sim1.5$~fm. As the wave function spreads outward, the density in the small-$r$ region is reduced, leading to a weaker effect of the Coulomb interaction.

% conclusion
From these observations, we conclude that the magnitude of the Coulomb-induced shift for the excited state is determined by a competition between the enhancement of the density near the origin due to the nodal wave function and the spread of the wave function caused by small binding energies. When the former effect dominates, one finds $|\Delta B_{1}|>|\Delta B_{2}|$, while for excited states located close to the threshold, where the latter effect becomes dominant, the opposite relation $|\Delta B_{1}|<|\Delta B_{2}|$ holds.

%==========================
\subsection{Nature of weakly bound states}
\label{subsec:weakbound}

\begin{figure}[tb]
\centering
\includegraphics[width=0.4\textwidth,bb=0 0 550 380]{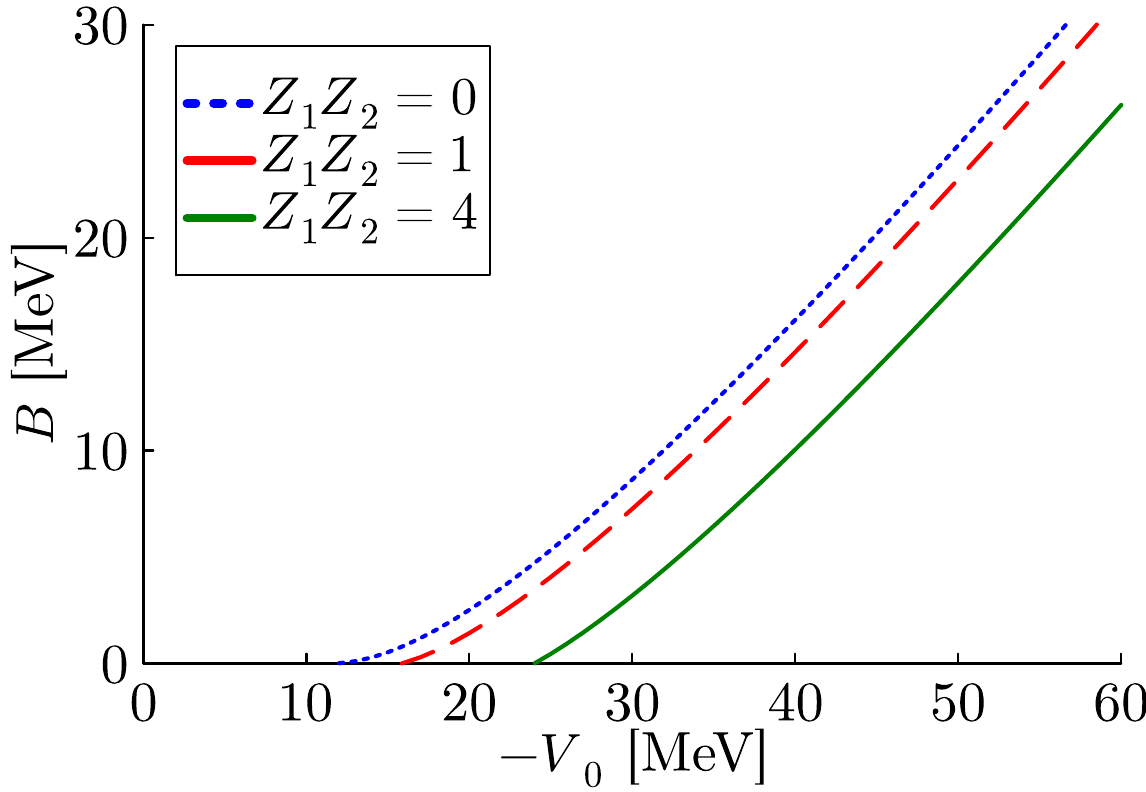}
\caption{Dependence of the binding energy $B$ on the well depth $-V_{0}$ for $Z_{1}Z_{2}=0$ (dotted line), $+1$ (dashed line), and $+4$ (solid line).}
\label{fig:V0B}
\end{figure}

% mass scaling near threshold
To discuss the properties of weakly bound states near the threshold, Fig.~\ref{fig:V0B} shows the dependence of the ground state binding energy $B$ on the depth of the potential well $|V_{0}|$ for $Z_{1}Z_{2}=0,+1,+4$. In all cases, when $|V_{0}|$ is sufficiently large, the binding energy increases approximately linearly with $|V_{0}|$. As the binding energy becomes smaller, however, the variation of $B$ with respect to $V_{0}$ becomes gradually weaker. It is known for the case with only a short-range interaction ($Z_{1}Z_{2}=0$) that the linear term vanishes in the limit $B\to0$  and the binding energy exhibits a quadratic behavior, $B\propto V_{0}^{2}$~\cite{Hyodo:2014bda}. In contrast, when the Coulomb repulsion is present, we find that the binding energy approaches zero while maintaining a finite slope, namely $B\propto V_{0}$ even in the threshold limit. This result indicates that the Coulomb repulsion qualitatively modifies the near-threshold mass scaling of weakly bound states.

\begin{figure}[tb]
\centering
\includegraphics[width=0.4\textwidth,bb=0 0 600 400]{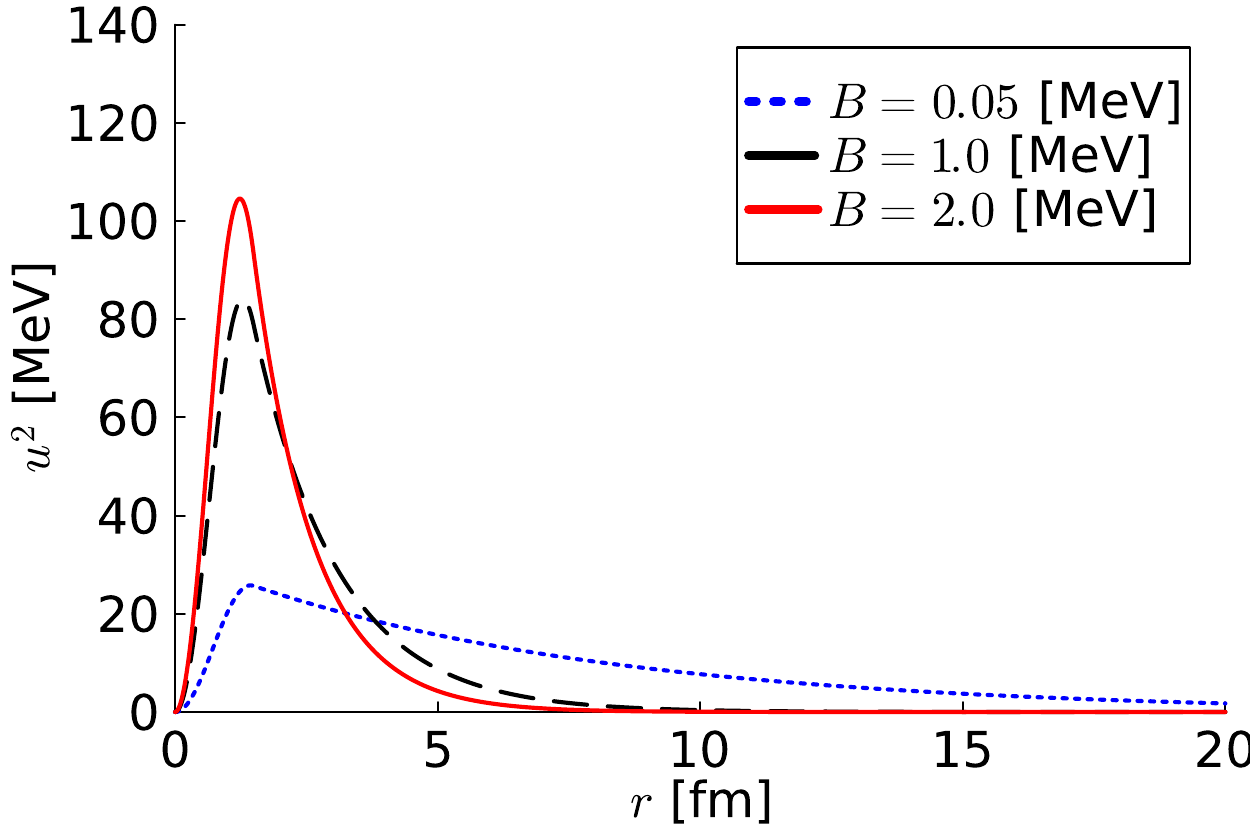}
\includegraphics[width=0.4\textwidth,bb=0 0 600 400]{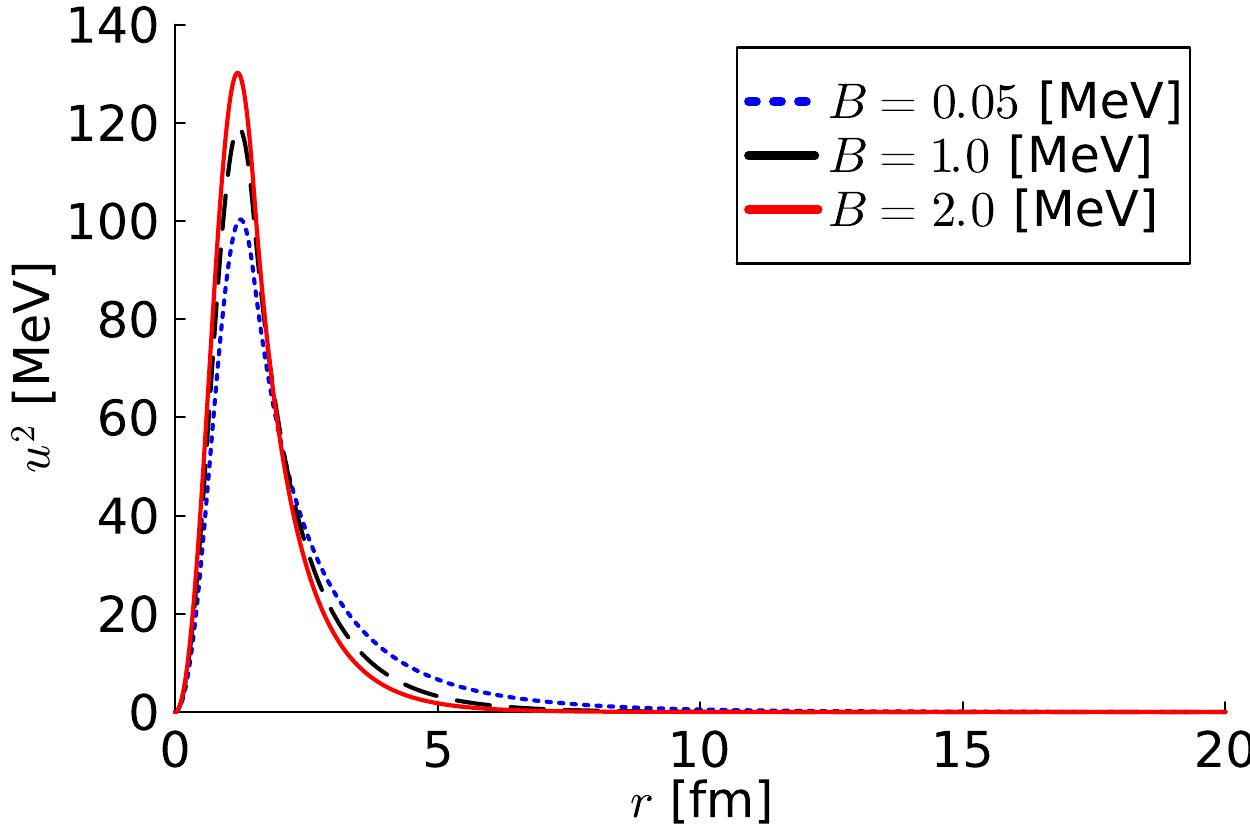}
\caption{Density distributions of weakly bound states with binding energies $B = 2.0$, $1.0$, and $0.05$~MeV. 
Left: $Z_{1}Z_{2}=0$. Right: $Z_{1}Z_{2}=+4$.}
\label{fig:wbwf}
\end{figure}

% relation to wave function
The modification of the mass scaling is closely related to the internal structure of the bound state in the limit $B \to 0$. Figure~\ref{fig:wbwf} shows the wave functions of bound states with binding energies $B = 2.0$, $1.0$, and $0.05$~MeV. The left panel corresponds to the case with only the short-range interaction, $Z_{1}Z_{2}=0$. As the binding energy decreases, the wave function extends farther into the asymptotic region. In the limit $B \to 0$, it is known that the wave function spreads to infinity, which leads to the emergence of low-energy universality~\cite{Braaten:2004rn,Naidon:2016dpf}. The right panel of Fig.~\ref{fig:wbwf} shows the wave functions of weakly bound states in the presence of the Coulomb repulsion with $Z_{1}Z_{2}=+4$. In this case, even in the limit $B \to 0$, the wave function does not extend to infinity but instead remains spatially localized within a finite range. This behavior can be interpreted as a consequence of the Coulomb repulsion which suppresses the wave function at large distances. The difference in the asymptotic behavior of the wave functions is directly related to the compositeness of the bound state~\cite{Hyodo:2013nka,Kinugawa:2024crb} in the $B \to 0$ limit. For systems governed solely by short-range interactions, the compositeness $X$ approaches unity as $B \to 0$, indicating a purely composite state~\cite{Hyodo:2014bda}. In contrast, it has been shown that in the presence of the Coulomb repulsion, the compositeness does not approach unity even in the $B \to 0$ limit~\cite{Kinugawa:2025kqr}.

% fate of bound state
It is also instructive to examine the fate of the bound state beyond the $B\to 0$ limit. 
For $s$-wave bound states generated by short-range interactions, the quadratic behavior $B \propto (V_{0})^{2}$ is associated with the transition of the bound state into a virtual state~\cite{Hyodo:2014bda}. In contrast, for states with finite angular momentum, the binding energy exhibits a linear dependence $B \propto V_{0}$, and the bound state evolves directly into a resonant state as it disappears. As observed in Fig.~\ref{fig:V0B}, the binding energy shows a linear behavior $B \propto V_{0}$ when the Coulomb interaction is included. This can be understood as a consequence of the Coulomb repulsion playing a role analogous to the centrifugal barrier in higher partial waves. Indeed, it is known that in systems with Coulomb repulsion, a bound state does not turn into a virtual state but instead undergoes a direct transition into a resonant state~\cite{Mochizuki:2024dbf,Kinugawa:2025kqr}.

%%%%%%%%%%%%%%%%%%%%
\section{Summary}
\label{sec:summary}
%%%%%%%%%%%%%%%%%%%%

In this work, we have studied the effect of the Coulomb force on bound states formed by short-range interactions. We have shown that the magnitude of the Coulomb-induced shift in the binding energy is governed by the spatial structure of the bound-state wave function. Furthermore, we demonstrated that the near-threshold mass scaling of bound states is qualitatively modified by the Coulomb interaction. In particular, in systems with Coulomb repulsion, the wave function of a weakly bound state remains localized with a finite spatial extent even in the limit of vanishing binding energy.

%
%\begin{table}[tbh]
%\caption{Captions to tables and figures should be sentences.}
%\label{t1}
%\begin{tabular}{ll}
%\hline
%AAA & BBB \\
%CCC & DDD \\
%\hline
%\end{tabular}
%\end{table}

%\begin{figure}[tbh]
%\includegraphics{fig01.eps}
%\caption{You can embed figures using the \texttt{\textbackslash includegraphics} command. Basically, figures should appear where they are cited in the text. You do not need to separate figures from the main text when you use \LaTeX\ for preparing your manuscript.}
%\label{f1}
%\end{figure}

%\bibliographystyle{h-physrev3}
%\bibliography{refs.bib}

%\begin{thebibliography}{9}
%\bibitem{cp} The abbreviation for JPS Conference Proceedings should be ``JPS Conf. Proc." in the reference list.
%\bibitem{jpsj} The abbreviation for the Journal of the Physical Society of Japan should be ``J. Phys. Soc. Jpn." in the reference list.
%\bibitem{ptep} The abbreviation for the Progress of Theoretical and Experimental Physics should be ``Prog. Theor. Exp. Phys." in the reference list.
%\bibitem{instructions} More abbreviations of journal titles are listed in ``Instructions for Preparation of Manuscript", which is available at our Web site (http://jpsj.jps.or.jp).
%\bibitem{format} F. Author, S. Author, and T. Author, Abbreviated journal title \textbf{volume in bold face}, initial page or article number (year of publication).
%\end{thebibliography}

\end{document}